\documentclass[a4paper,preprintnumbers,floatfix,superscriptaddress,prl,twocolumn,showpacs,notitlepage,aps,10pt]{revtex4-2}
\usepackage[sc,osf]{mathpazo}
\usepackage{amsmath,amsfonts,amssymb}
\usepackage{graphicx}
\usepackage{xcolor}
\usepackage{physics}
\usepackage{mathtools}
\usepackage[breaklinks=true]{hyperref}
\hypersetup{
  colorlinks   = true,
  urlcolor     = blue,
  linkcolor    = blue,
  citecolor   = red
}
\usepackage{cleveref} 
\crefformat{equation}{Eq.~(#2#1#3)}
\crefformat{figure}{Figure~#2#1#3}
\crefformat{table}{Table~#2#1#3}

\newcommand{\cnot}{\textsc{cnot}}
\newcommand{\swap}{\textsc{swap}}
\newcommand{\cz}{\textsc{cz}}

\begin{document}

\title{Quantifying Quantum Causal Influences}

\author{Lucas Hutter}
\affiliation{Institute of Physics, Federal University Fluminense, Niterói, Brazil}

\author{Rafael Chaves}
\affiliation{International Institute of Physics, Federal University of Rio Grande do Norte, 59070-405 Natal, Brazil}
\affiliation{School of Science and Technology, Federal University of Rio Grande do Norte, 59078-970 Natal, Brazil}

\author{Ranieri Vieira Nery}
\affiliation{International Institute of Physics, Federal University of Rio Grande do Norte, 59070-405 Natal, Brazil}

\author{George Moreno}
\affiliation{International Institute of Physics, Federal University of Rio Grande do Norte, 59070-405 Natal, Brazil}
\affiliation{Departamento de Computação, Universidade Federal Rural de Pernambuco, 52171-900, Recife, Pernambuco, Brazil}

\author{Daniel Jost Brod}
\affiliation{Institute of Physics, Federal University Fluminense, Niterói, Brazil}

\begin{abstract}
Causal influences are at the core of any empirical science, the reason why its quantification is of paramount relevance for the mathematical theory of causality and applications. Quantum correlations, however, challenge our notion of cause and effect, implying that tools and concepts developed over the years having in mind a classical world, have to be reevaluated in the presence of quantum effects. Here, we propose the quantum version of the most common causality quantifier, the average causal effect (ACE),  measuring how much a target quantum system is changed by interventions on its presumed cause. Not only it offers an innate manner to quantify causation in two-qubit gates but also in alternative quantum computation models such as the measurement-based version, suggesting that causality can be used as a proxy for optimizing quantum algorithms. Considering quantum teleportation, we show that any pure entangled state offers an advantage in terms of causal effects as compared to separable states. This broadness of different uses showcases that, just as in the classical case, the quantification of causal influence has foundational and applied consequences and can lead to a yet totally unexplored tool for quantum information science.
\end{abstract}

\maketitle

In spite of the mantra in statistics that "correlation does not imply causation", a central goal of any quantitative science is precisely that: to infer cause and effect relations from observed data \cite{pearl2009causality,spirtes2000causation}. In fact, as stated by Reichenbach's principle \cite{reichenbach1991direction}, correlations between two events do imply some causation. Either one has a direct influence over the other or a third event acts as a common cause for both of them. 
Within this context, given variables $A$ and $B$, the basic aim of causal inference is to distinguish how much of their observed correlations are due to direct causal influences, rather than due to a potential common cause $\Lambda$. However, if we do not have empirical access to $\Lambda$, which is then treated as a latent/hidden variable, both models---common cause or direct causal influences---generate the same set of possible correlations that cannot be set apart from observations alone. With that aim, one has to rely on interventions \cite{pearl2009causality}. By intervening on $A$, we fix it to a value independent of $\Lambda$ such that any remaining correlations between $A$ and $B$ can unambiguously be traced back to a direct influence $A \rightarrow B$, a fundamental result with a wide range of applications \cite{pearl2009causal,friedman2004inferring,peters2017elements,angrist1996identification,glymour2001mind,shipley2016cause}.

Notwithstanding all the successes of causality theory, since Bell's theorem \cite{bell1964einstein} it is known that the classical notions of cause and effect break down at the quantum level. 
Not only the notion of a causal structure has to be generalized in order to include quantum states \cite{henson2014theory,fitzsimons2015quantum,ried2015quantum,fritz2016beyond,chaves2015information,costa2016quantum,allen2017quantum,aaberg2020semidefinite} or the possibility of superposition of causal orders \cite{oreshkov2012quantum,barrett2019quantum,barrett2021cyclic}; but also the meaning and applicability of Bell inequalities as a causal compatibility tool \cite{pearl1995testability} have to be reevaluated \cite{chaves2018quantum}, and tests employed to bound the causal effects \cite{balke1997bounds} have to be readjusted \cite{gachechiladze2020quantifying}.
Given that, a fundamental question reemerges: how can we quantify quantum causal effects? Complementary frameworks for reasoning about quantum causal influences have been developed \cite{ried2015quantum,chaves2015information,allen2017quantum,barrett2019quantum,barrett2021cyclic} and explicit quantifiers of causality have already been proposed \cite{escola2021quantifying,Yiquantum2022}. Nevertheless, the quantum generalization of the most widely used and intuitive quantifier of causality in the classical case, the so-called average causal effect (ACE) \cite{pearl2009causality,angrist1996identification,balke1997bounds,janzing2013quantifying,gachechiladze2020quantifying,miklin2022causal}, has not yet been achieved. That is the main goal of this Letter.


Using the trace distance, we propose a quantum version of the ACE. It quantifies the causal influence that an intervention on a system might have on a resulting quantum state. We show the applicability of our framework in a number of paradigmatic quantum information scenarios. We start quantifying causal influences in two-qubit gates and discussing the advantages of our approach as compared to other recent proposals \cite{escola2021quantifying}. Within the context of measurement-based quantum computation \cite{raussendorf2001one,briegel2009measurement} and quantum teleportation \cite{bennett1993teleporting}, we show that separable states imply a limited amount of causal influence, a restraint that can be surpassed by any pure entangled state. Thus, our quantum causality quantifier not only provides a natural extension of a widely used and acknowledged classical tool but also can be seen as a novel witness of non-classical behavior.



\textit{Quantum Average Causal Effect -- }
Suppose we observe correlations between variables $A$ and $B$, that is, their probability distribution does not factorize as $p(a,b)\neq p(a)p(b)$. From Reichenbach's principle \cite{reichenbach1991direction}, the most general causal model explaining such correlations might involve direct influences as well as a common cause $\Lambda$ that, for a variety of reasons, might not be directly observed. Thus, at least in a classical description, the conditional observational distribution $p(b \vert a)$ can be decomposed as
\begin{equation}
    p(b\vert a)=\sum_{\lambda}p(\lambda\vert a)p(b\vert a,\lambda).
\end{equation}
If, however, an intervention is performed on $A$, an operation denoted as $do(a)$, the interventional distribution is now
\begin{equation}
    p(b\vert do(a))=\sum_{\lambda}p(\lambda)p(b\vert a,\lambda),
\end{equation}
where $p(b|do(a))$ denotes the probability of $b$ when variable $A$ is set by force to be $a$, that is, any potential influence it might have had from the common cause $\Lambda$ is erased. Importantly, interventions bring in a natural way for quantifying causality. For instance, if $a$ and $b$ are binary variables, a widely used quantifier of the causal influence from $A$ to $B$, the \textit{average causal effect} (ACE) \cite{pearl2009causality,angrist1996identification,balke1997bounds,janzing2013quantifying,gachechiladze2020quantifying,miklin2022causal}, is defined  as
\begin{equation}
ACE= \lvert P(b_1| do(a_1))- P(b_1| do(a_0)) \rvert, \label{eq:ACE_Pearl}
\end{equation}
measuring the change in the distribution $p(b_1)=p(b=1)$ of the variable $B$ depending whether the value of $A$ is set to $a=1$ or $a=0$. Notice that
\begin{equation*}
P(b_1| do(a_1))- P(b_1| do(a_0))=P(b_0| do(a_0))-P(b_0| d
o(a_1)),
\end{equation*}
therefore \cref{eq:ACE_Pearl} accounts for the influence $A$ on the full probability distribution of values of $B$. In contrast, when $A$ and $B$ can assume more than two values, generalizations of \cref{eq:ACE_Pearl} are not unique. 




For simplicity, first assume that only $B$ can have more than two values. If we want to measure the largest causal influence $A$ has over $B$, a natural generalization is to maximize the right-hand side of \cref{eq:ACE_Pearl} over all values of $b$ \cite{gachechiladze2020quantifying}, such that
\begin{equation} \label{eq:ACERChaves}
ACE_{max}=\max_{b} \abs{P(b|do(a_1))-P(b|do(a_0))}.
\end{equation}
This definition, however, might not capture the full causal influence from $A$ to $B$, if that influence is very spread through the event space of $B$. To illustrate, suppose that $B$ can assume integer values from $1$ to $2N$. If $a=0$ (resp. $a=1$), the probability distribution over $B$ is the uniform distribution over integers from 1 to $N$ (resp.\ $N+1$ to $2N$). The ACE, as defined by \cref{eq:ACERChaves}, decreases as $N$ increases. And yet, changing the value of $A$ clearly has a large effect on the distribution of $B$. This example illustrates the extent to which $ACE_{max}$ is sensitive to a coarse-graining of the probability distribution. Since our intention is to quantify causal influence in quantum protocols, it makes sense to allow for arbitrary coarse-grainings on outcomes of quantum measurements---after all, we can always encode a coarse-graining strategy as degeneracies in the measured observable. 

Building on that, we propose a generalization of \cref{eq:ACE_Pearl} 
based on the well-know \textit{total variation distance} (TVD), the largest possible difference that the two distributions can assign to the same event, given by
\begin{equation}
\delta(P,Q)=\frac{1}{2}\sum_{x\in X}\abs{P(x)-Q(x)},
\end{equation}
where $P$ and $Q$ are two probability distributions over $X$. The ACE can then be defined as
\begin{equation}
ACE_{TVD}=\frac{1}{2}\sum_{b}\abs{P(b|do(a_1))-P(b|do(a_0))}.\label{eq:ACETVD}
\end{equation}
While it reduces to \cref{eq:ACE_Pearl} when $B$ is binary, it actually returns the largest value of $ACE_{max}$ over all possible coarse-grainings of the distribution of $B$.

To generalize \cref{eq:ACETVD} for a quantum system, there are two choices. The first is to suppose we have some set of possible measurement bases and compute the $ACE_{TVD}$ at the level of the probability distribution over measurement outcomes in these bases. Often, this is desirable, since it operates directly at the level of outcomes \cite{chaves2018quantum,gachechiladze2020quantifying,agresti2020experimental}---the success probability of a quantum game or protocol might be stated in terms of these quantities, as typically done within device-independent quantum information \cite{scarani2012device}. However, there are in principle infinitely many choices of measurement bases, and different protocols or setups can differ on how much information the experimenter or measuring agent has over which bases they should measure. Therefore, it can also be meaningful to measure directly the causal influence of a parent variable on the resulting quantum state, agnostic to which basis (if any) it will be measured in.

Following this reasoning, we propose a generalization of the ACE for quantum states in terms of the trace distance (TD), a well-known generalization of the TVD measuring the distance between two density matrices $\rho$ and $\sigma$, defined as
\begin{equation}
TD(\rho,\sigma):= \frac{1}{2} \Tr\left(\sqrt{(\rho-\sigma)^2}\right)= \frac{1}{2} \sum_i \abs{\lambda_i},
\end{equation}
where $\lambda_i$ are the eigenvalues of the matrix $(\rho-\sigma)$. Just like the TVD accounts for all classical ``strategies'' (i.e.\ choices of coarse-graining), the TD accounts for all possible quantum strategies. More concretely, the trace distance between two states corresponds to the maximum TVD between the two probability distributions that would arise from measuring those states with the same POVM.

If $A$ is a classical binary variable, then the quantum ACE is naturally defined as
\begin{equation} \label{eq:ACETD}
ACE_Q =  TD(\rho_B(do(a_1)),\rho_B(do(a_0)),
\end{equation}
where $\rho_B(do(a_0))$ is the density matrix that describes the state at $B$ given the intervention $do(a_0)$. In many cases, however, and particularly for the applications we consider later on, $A$ actually corresponds to some pure (qubit) quantum state. More concretely, $A$ could  correspond to 
any state in the Bloch sphere, and so \cref{eq:ACETD} is no longer well defined. We thus generalize it as follows
\begin{equation} \label{eq:ACEQ}
    ACE_{Q} = \underset{a_0,a_1}{\mathbb{E}} TD(\rho_B(do(a_1)),\rho_B(do(a_0)),
\end{equation}
where we now average over the choice of $a_0$ and $a_1$. Following \cite{ried2015quantum}, do-interventions on quantum states are obtained simply by tracing whatever state represents $A$ and replacing it with a pure state, and subsequently averaging over all possible states of $A$. Clearly, which average must be performed depends on the nature of the variable $A$. For instance, if it is an arbitrary state in the Bloch sphere, the natural choice is the uniform (Haar) distribution \cite{meckes_2019,Mezzadri_2007}.

\begin{figure}[t]
    \centering
    \includegraphics[width=0.4\textwidth]{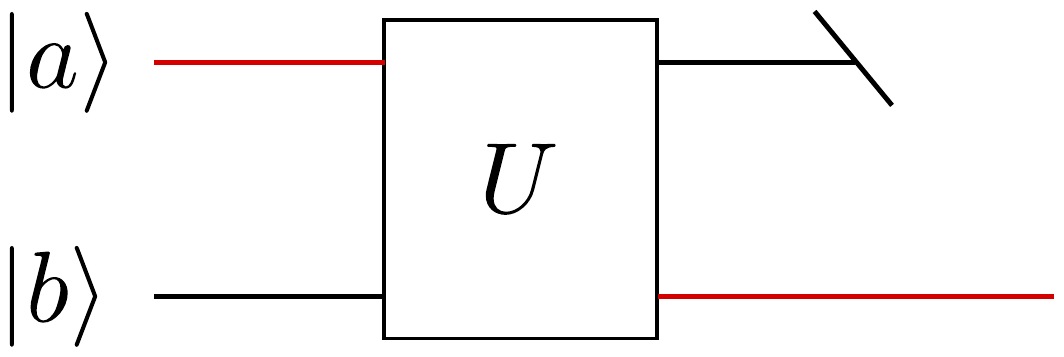}
    \caption{The setup used to calculate the causal influence of quantum gates. Here the influence is measured from the entry $\ket{a}$ to the output of $\ket{b}$ (red lines). We perform a Haar-random average over $\ket{b}$ and a partial trace over the first output qubit.}
    \label{fig:gatescircuits}
\end{figure}

\textit{Causal influences in two-qubit gates-- }
We consider a two-qubit gate, $U$, acting on a pair of qubits labelled $A$ and $B$. We wish to compute the $ACE_Q$ from the input state of qubit $A$ onto the output of qubit $B$. We consider that this gate might be embedded into a larger quantum circuit, but that we perform a do-intervention  on qubit $A$, replacing it by some pure state $\ket{A}$ \cite{ried2015quantum}. As there is no reason for $B$ to be diagonal in any particular basis, we perform a Haar-random average over the input of $B$. As we are also not interested in the output of qubit $A$, it is traced out after the application of $U$. The entire procedure, shown in \cref{fig:gatescircuits}, can be summarized by
\begin{equation} \label{eq:ACEU}
    ACE_{Q}(U) = \underset{\ket{a}}{\mathbb{E}} \underset{\ket{b}}{\mathbb{E}} TD(\rho(b|do(a),\rho(b|do(a^\perp)),
\end{equation}
where we used the shorthand
\begin{equation}
    \rho(b|do(a)) = \textrm{tr}_A\left(U \ketbra{a,b}{a,b} U^{\dagger}\right).
\end{equation}


The average over choices of intervention is done as follows. First, we choose some state $\ket{a}$, and assume the intervention consisted of choosing either of the antipodal states in the Bloch sphere $\ket{a}$ and $\ket{a^\perp}$. We then average the result uniformly over $\ket{a}$. We could have chosen to average uniformly over two independent choices of states $\ket{a_0}$ and $\ket{a_1}$. However, this is computationally more costly and seems to lead simply to a reduction by a constant multiplicative factor. It is also intuitive that, given any state $\ket{a}$, the largest influence over $B$ is obtained by choosing between either $\ket{a}$ and $\ket{a^\perp}$.

 \begin{table}[ht]
     \centering
     \begin{tabular}{|c||c|}
        \hline
        Gate & $ACE_Q$ \\
        \hline\hline
        $\textrm{Local}$ & $0$ \\
        \hline
        $\cnot$ & $\pi/8$ \\
        \hline
        $\cz$ & $\pi/8$  \\
        \hline
        $B$ gate & 0.5878 \\
        \hline
        $\sqrt{\swap}$ & 0.6427 \\
        \hline
        $\swap$ & 1 \\
        \hline
     \end{tabular}
     \caption{$ACE_Q(U)$ for some noteworthy two-qubit gates. As expected, the causal influence on a $\cnot$ gate is the same in both directions. The $B$ gate was defined in \cite{zhang2004minimum}, and is an optimal two-qubit gate in the sense that any other gate can be decomposed using only two copies of it (compared to three $\cnot$ gates). This distinction manifests in the fact that $ACE_Q$ is much larger for the $B$ gate than for the $\cnot$
     }
     \label{tab:results}
 \end{table}

Our results, for a few paradigmatic quantum gates, are summarized in \cref{tab:results} and detailed in the Supplemental Material \cite{SM}. It is natural that the $\swap$ gate displays the largest possible value of causal influence: if the states of qubits $A$ and $B$ get swapped, then $A$ has maximal influence over $B$ irrespective of anything else. Another virtue of our definition is that it is basis invariant. As a consequence, consider the \cnot~gate: it flips the target qubit if the control qubit is in the $\ket{1}$ state, and does nothing otherwise. Thus, one can imagine that the influence only exists from the (input) control qubit onto the (output) target qubit, or at least that it is stronger in that direction. Our measure, however, attributes the same causal influence from the control to the target in a \cnot~gate as vice-versa, which is to be expected since these roles can be flipped by a local change of basis. Finally, our definition has a natural scale, ranging from 0 (for local gates) to 1 (for the $\swap$ gate). Thus, not only our causality measure has a fundamental motivation since it is a generalization of the well-known ACE \cite{pearl2009causality}, it also displays a number of advantages that can be showcased by comparison with another recent proposal \cite{escola2021quantifying}. There, the \cnot~gate does not have the same value of causal influence in both directions, and neither does their definition has a natural scale, which is inferred by averaging over Haar-random unitary gates.

\textit{One-way model of quantum computation-- }In the measurement-based quantum computation (MBQC) model \cite{briegel2009measurement}, the interactions between the qubits and unitary operations required to execute a given algorithm are replaced by the initial entanglement in a graph-state \cite{hein2006entanglement} and the possibility of performing local adaptative measurements. Measurements in the computational basis $\left\{ \ket{0},\ket{1} \right\}$ disconnect unnecessary qubits from the graph-state while measurements on the $X-Y$ plane of the Bloch-sphere, represented by the eigenstates  $\ket{a}=(1/\sqrt{2})(\ket{0}+e^{i \phi_a} \ket{1})$ and $\ket{a^\perp}=(1/\sqrt{2})(\ket{0}-e^{i \phi_a} \ket{1})$, perform the desired quantum gates. Quantum computation is then characterized by a collection of angles defining the measurement basis for each qubit, as well as a list of dependencies of these angles on outcomes of previous measurements. There is a feed-forward of classical information (measurement outcomes) along the computation, explaining why this approach is also known as the one-way model \cite{raussendorf2001one}.

A building block for MBQC is a two-qubit graph state $\ket{G_2}=(1/\sqrt{2})(\ket{0+}+\ket{1-}$). One measures the first qubit in the basis $\left\{ \ket{a},\ket{a^\perp} \right\}$, obtaining outcome $s=0,1$. The second qubit is then projected to $X^s R_x(\phi_a)\ket{0}$, where $R_x(\phi_a)=e^{-i\phi_a X/2}$ and $X^s$ is the so-called by-product of the computation.  If $s=0$ (i.e.\ outcome $\ket{a}$) then the desired rotation $R_x(\phi_a)$ was achieved. Otherwise, if the outcome was $s=1$ (i.e.\ outcome $\ket{a^\perp}$) one has to correct the extra $X$ term. By concatenating two-qubit graph states we can perform arbitrary single-qubit gates as well as a CNOT gate, and thus universal quantum computation.

Our aim is to investigate how the causal influence from $A$ to $B$ behaves in this MBQC building block, that is, the influence of the measurement basis (defining the desired gate) on the state that is actually prepared on the remaining qubit, particularly when we consider that state $\ket{G_2}$ is replaced by some imperfect alternative $\rho_{in}$. In this case, $ACE_Q$ is
\begin{equation}
    ACE_{Q}(\rho_{in}) = \underset{\ket{a}}{\mathbb{E}} TD(\rho_B(do(a)),\rho_B(do(a^\perp)), \label{eq:ACE_MBQC}
\end{equation}
where $\rho_B(do(a))$ is the output state when the first qubit is measured in the $\left\{ \ket{a},\ket{a^\perp} \right\}$ basis and the resource state is $\rho_{in}$. When $\rho_{in} =\ketbra{G_2}{G_2}$ the basis choice perfectly defines the output state, and hence $ACE_Q(\ketbra{G_2}{G_2})=1$, as expected.

\begin{figure}[ht]
    \centering
    \includegraphics[width=0.5\textwidth]{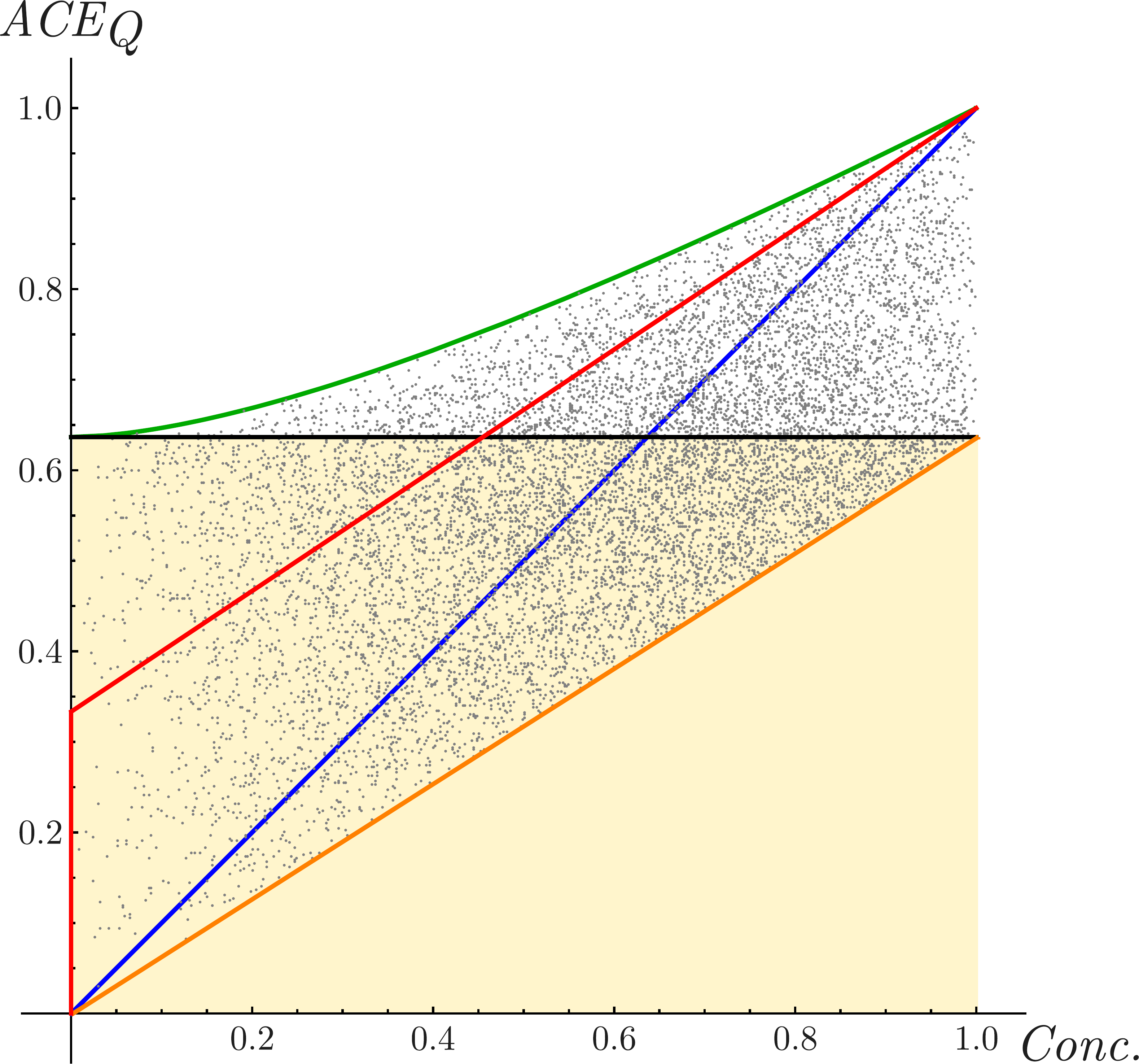}
    \caption{$ACE_Q$ as a function of the concurrence. The green, red, blue, and orange curves correspond to families of states $\ket{H^{\epsilon}_2}$, $\rho_{iso}$, $\ket{G^{\epsilon}_2}$, and $\ket{F^{\epsilon}_2}$, respectively, as defined in the main text. The black horizontal line delimits the shaded region and corresponds to $ACE^{sep}_Q = 2/\pi$. Points correspond to 10000 uniformly-sampled pure states.}
    \label{figure:RSP}
\end{figure}

As proven in the Supplemental Material \cite{SM}, if the resource state is separable, that is, $\rho_{in}=\rho_{sep}=\sum_{i}p_i \rho^i_A \otimes \rho^i_B$, then  $ACE_{Q}(\rho_{sep}) \leq 2/\pi$, with equality achieved for state $\ket{0+}$. In turn, for a pure entangled state $\ket{G^{\epsilon}_2}= \sqrt{\epsilon}\ket{0+}+\sqrt{1-\epsilon}\ket{1-}$, we have that $ACE_Q(\ketbra{G^{\epsilon}_2}{G^{\epsilon}_2})= \frac{2}{\pi} E[(1-2\epsilon)]$, where $E(k)$ is the complete elliptic integral of the second kind (see \cref{figure:RSP} and \cite{SM}). That is, up to local unitaries, any pure entangled state surpasses the maximum quantum ACE achievable by separable states, which can be seen as a sort of advantage in the one-way model. 

In \cref{figure:RSP} we show the relation between the concurrence \cite{Hill1997conc} of two-qubit states and their quantum ACE when used as a resource in the one-way model. The figure shows uniformly sampled (pure) quantum states, as well as curves corresponding to specific parameterized families of states, such as pure partially-entangled states $\ket{G^{\epsilon}_2}$, $\ket{F^{\epsilon}_2}= (H\otimes \openone) \ket{G^{\epsilon}_2}$, and $\ket{H^{\epsilon}_2}= (H\otimes H) \ket{G^{\epsilon}_2}$, as well as the depolarized state $\rho_{iso}=\epsilon\ketbra{G_2}{G_2}+(1-\epsilon)\openone/4$. The shaded region is delimited by the highest value achieved by a separable state, $ACE^{sep}_Q = 2/\pi$. Clearly, for a given concurrence, states $\ket{G^{\epsilon}_2}$ and $\ket{F^{\epsilon}_2}$ serve as upper and lower bounds on the $ACE_Q$, respectively. For more details, see the Supplemental Material \cite{SM}.

\textit{Quantum Teleportation --} The final scenario we analyze from the perspective of causal influence is quantum teleportation \cite{bennett1993teleporting}. We consider the standard teleportation protocol, where Alice wants to teleport some state $\ket{a}$ to Bob, and they share a Bell pair. Alice applies a Bell basis measurement on $\ket{a}$ together with her end of the Bell pair and informs Bob of the outcome. He finally applies a quantum gate (which depends on Alice`s outcome) to his end of the Bell pair, thereby recovering state $\ket{a}$. 

The $ACE$ we consider in the teleportation scenario is defined analogously as in \cref{eq:ACE_MBQC}, where $\rho_B(do(a))$ is the output state in Bob`s side of the protocol when Alice prepares one of two orthogonal states $\left\{ \ket{a},\ket{a^\perp} \right\}$, and where we assume they follow the teleportation protocol exactly. As before, $\rho_{in}$ is some imperfect entangled state that will replace their initial Bell pair. If $\rho_{in} = \ketbra{F_2^{1/2}}{F_2^{1/2}}$, where $\ket{F^{\epsilon}_2}= \sqrt{\epsilon}\ket{00}+\sqrt{1-\epsilon}\ket{11}$, then the teleportation is successful and $ACE_Q = 1$.

The qualitative behavior of $ACE_Q$ in the case of teleportation is similar to that of \cref{figure:RSP}. This is not surprising, since measurement-based quantum computing is inspired by a scheme that uses teleportation as a computational primitive \cite{Gottesman_1999}. Any entangled state can exhibit an $ACE_Q$ better than the best separable strategy (where now $ACE_{Q}(\rho_{sep}) = 1/2$). One main difference is that there is no nontrivial lower bound in the case of teleportation, i.e., as we observe numerically, for any given concurrence there exists some state which has a $ACE_Q$ of zero. The upper bound is achieved by $\ket{F^{\epsilon}_2}$. The plots and a more thorough analysis can be found in the Supplemental Material \cite{SM}.

\textit{Discussion --} Quantifying causal influences with the use of interventions is a central concept and tool for causal inference, with applications ranging from the reconstruction of genetic networks \cite{friedman2004inferring} to social studies \cite{shipley2016cause} and learning algorithms \cite{peters2017elements}. Given that quantum theory is at odds with the classical notion of causality, it is natural to seek a generalization of the most common causality quantifier, the average causal effect (ACE), and employ it to analyze paradigmatic quantum information processing protocols. Here we propose a quantum ACE based on the trace distance, quantifying how much a target quantum system is changed by interventions on its presumed cause. 

Our approach offers an innate manner to quantify causation in two-qubit gates, with a natural scale that ranges from $ACE_Q=0$ for local gates up to $ACE_Q=1$ for a $\swap$ gate. Interestingly, the $CNOT$ gate, three of which are required to perform any other two-qubit gate, has $ACE_Q=\pi/8$. In turn, the $B$ gate \cite{zhang2004minimum}, two of which are sufficient to compose any other two-qubit gate, has $ACE_Q\approx 0.5878$. This suggests that quantifiers of causality can be used as a proxy for optimizing quantum circuits. We also obtain results for an alternative quantum computation model, based on measurements \cite{briegel2009measurement}, showing that, for its two-qubit building block, any pure entangled state offers an advantage in terms of $ACE_Q$ as compared to separable states. A similar result is valid for quantum teleportation, pointing out that our quantifier of quantum causality can be employed as a witness of non-classicality in a wide range of information processing scenarios. This broadness of different uses shows that, just as in the classical case, the quantification of causal influence has foundational and applied consequences, a topic that deserves further investigation and for which our results might trigger further developments.

\begin{acknowledgments}
This work was supported by Instituto Nacional de Ciência e Tecnologia de Informação Quântica (INCT-IQ), CNPq (Grant No 307295/2020-6), CAPES, FAPERJ, the Serrapilheira Institute (Grant No. Serra-1708-15763), the Simons Foundation (Grant Number 884966, AF). The authors would like to thank D.\ Jonathan for fruitful discussions.
\end{acknowledgments}

\bibliography{bibliography}

\onecolumngrid

\appendix

\section{Causal influence of two-qubit gates - Details}

As discussed in the main text, we define our measure of causal influence of two-qubit gates as follows. We initialize both input qubits in arbitrary states, which we parameterize as:
\begin{align}
\ket{a}=&\cos{\left(\tfrac{\theta_1}{2}\right)}\ket{0}+e^{i \phi_1}\sin{\left(\tfrac{\theta_1}{2}\right)}\ket{1}, \\
\ket{b}=&\cos{\left(\tfrac{\theta_2}{2}\right)}\ket{0}+e^{i \phi_2}\sin{\left(\tfrac{\theta_2}{2}\right)}\ket{1}.
\end{align}
In this parameterization, we can also write
\begin{equation}
\lvert{a^\perp}\rangle=\sin{\left(\tfrac{\theta_1}{2}\right)}\ket{0}-e^{i \phi_1}\cos{\left(\tfrac{\theta_1}{2}\right)}\ket{1}.
\end{equation}
We can now rewrite Eq.\ (10) of the main text as
\begin{equation}
ACE_{Q}(U) = \underset{\ket{a}}{\mathbb{E}} \underset{\ket{b}}{\mathbb{E}} TD[\rho(b|do(a),\rho(b|do(a^\perp))],
\end{equation}
where we used the shorthand
\begin{equation} \label{twoqubit_state}
\rho(b|do(a)) = \textrm{tr}_A\left(U (\ketbra{a}{a}\otimes\ketbra{b}{b}) U^{\dagger}\right),
\end{equation}
and $TD$ is the trace distance between the two states. We can now perform the Haar-random averages over the inputs, $\underset{\ket{a}}{\mathbb{E}} \underset{\ket{b}}{\mathbb{E}}$, by integrating over the angles $\{\theta_1,\phi_1,\theta_2,\phi_2\}$. Recall that we chose the two intervention states in qubit $A$ as $\ket{a}$ and $\lvert{a^\perp}\rangle$, and then averaging only over the choice of $\ket{a}$. We could have chosen to average over two independent intervention states, but we verified numerically that this was computationally more expensive and only lead to a reduction of the $ACE_Q$ by a constant fraction.

The uniform average over the Bloch sphere can be obtained, for an arbitrary function $f(\theta,\phi)$, by performing the following integral
\begin{equation} \label{eq:Haar}
\langle f \rangle = \frac{1}{4\pi} \int_{0}^{2\pi} \int_{0}^{\pi} f(\theta,\phi) \sin{\theta} d\theta d\phi.
\end{equation}

We also can write explicitly the matrices used to evaluate Table I in the main text. First, the entry for ``Local'' simply means any matrix of the form
\begin{equation}
Q\otimes P,
\end{equation}
with $Q$ and $P$ being any two single qubit gates. Beyond local gates, we have:
\begin{align}
\cnot = \ketbra{0}{0}\otimes \openone + \ketbra{1}{1}\otimes X &= \begin{pmatrix}
1 & 0 & 0 & 0 \\
0 & 1 & 0 & 0 \\
0 & 0 & 0 & 1 \\
0 & 0 & 1 & 0
\end{pmatrix},
\end{align}

\begin{align}
B = \exp\left[i (\tfrac{\pi}{4} X\otimes X + \tfrac{\pi}{8} Y\otimes Y)\right] & =\begin{pmatrix}
\cos(\tfrac{\pi}{8}) & 0 & 0 & i \sin(\tfrac{\pi}{8}) \\
0 & \sin(\tfrac{\pi}{8}) & i \cos(\tfrac{\pi}{8}) & 0 \\
0 & i \cos(\tfrac{\pi}{8}) & \sin(\tfrac{\pi}{8}) & 0 \\
i \sin(\tfrac{\pi}{8}) & 0 & 0 & \cos(\tfrac{\pi}{8})
\end{pmatrix},
\end{align}

\begin{align}
\cz = \ketbra{0}{0}\otimes \openone + \ketbra{1}{1}\otimes Z &= \begin{pmatrix}
1 & 0 & 0 & 0 \\
0 & 1 & 0 & 0 \\
0 & 0 & 1 & 0 \\
0 & 0 & 0 & -1
\end{pmatrix},
\end{align}

\begin{align}
\sqrt{\swap} = \exp\left[i \tfrac{\pi}{8} ( X\otimes X + Y\otimes Y + Z\otimes Z)\right] &=\begin{pmatrix}
e^{i \pi/8} & 0 & 0 & 0 \\
0 & \tfrac{1}{\sqrt{2}} e^{-i \pi/8} & i \tfrac{1}{\sqrt{2}} e^{-i \pi/8} & 0 \\
0 & i \tfrac{1}{\sqrt{2}} e^{-i \pi/8} & \tfrac{1}{\sqrt{2}} e^{-i \pi/8} & 0 \\
0 & 0 & 0 & e^{i \pi/8}
\end{pmatrix},
\end{align}

\begin{align}
\swap= \frac{1}{2} (\openone \otimes \openone+ X\otimes X+ Y\otimes Y+ Z\otimes Z)&= \begin{pmatrix}
1 & 0 & 0 & 0 \\
0 & 0 & 1 & 0 \\
0 & 1 & 0 & 0 \\
0 & 0 & 0 & 1
\end{pmatrix}.
\end{align}

\section{Causal influence in the one-way model of quantum computation - Details}

As shown in the main text, the one-way quantum computer starts with the building block graph state
\begin{equation} \label{eq:graph}
\ket{G_2}=\frac{1}{\sqrt{2}} (\ket{0+}+\ket{1-}).
\end{equation}
Measurement are performed in the basis $\{\ket{a},\lvert{a^\perp}\rangle\}$ given by
\begin{align}
\ket{a}&= \frac{1}{\sqrt{2}}(\ket{0}+ e^{i\phi_a} \ket{1}), \\
\lvert{a^\perp}\rangle&= \frac{1}{\sqrt{2}}(\ket{0}- e^{i\phi_a} \ket{1}).
\end{align}
After the measurement of the first qubit, the second qubit collapses to
\begin{equation}
\ket{b}= X^s R_x(\phi_a) \ket{0},
\end{equation}
where $R_x(\phi)= e^{-i\phi X/2}$, and $s=0$ if outcome $\ket{a}$ was observed and $s=1$ otherwise. The $X^s$ gate is a conditional correction applied to the remaining qubit necessary for the protocol to succeed.

We can now compute the $ACE_Q$ as
\begin{equation}
ACE_Q(\rho_{in})= \underset{\ket{a}}{\mathbb{E}} TD(\rho_B(do(a)),\rho_B(do(a^\perp)),
\end{equation}
where labels $A$ and $B$ correspond to the first and second qubit. The density matrix $\rho_B(do(a))$ after the measurement and  correction is
\begin{equation} \label{rhoBdo(a)}
\rho_B(do(a))= \textrm{tr}_A\left(\Pi_{a}\rho_{in}\right) + X\left(\textrm{tr}_A\left(\Pi_{a^\perp} \rho_{in}\right) \right)X,
\end{equation}
$\Pi_{a}=\ketbra{a}{a}$ and $\Pi_{a^\perp}=(\openone-\Pi_{a})$ are projectors on the two measurement outcomes, and $\rho_{in}$ is the shared input state. For the protocol as described above, $\rho_{in} = \ketbra{G_2}{G_2}$, though we consider alternative shared states shortly.

In this case, since the two choices of intervention state lie on the equator of the Bloch sphere, we perform a uniform average over just that subspace. More concretely, for an arbitrary function $f$, we replace Eq.\ \eqref{eq:Haar} by
\begin{equation}
\langle f \rangle = \frac{1}{2\pi} \int_{0}^{2\pi} f(\phi) d\phi.
\end{equation}

We also characterize the causal influence in the one-way model when we change the shared resource state. More precisely, how much influence can we observe when the state is partially entangled, or only classically correlated? To that end, we replace the entangled state in Eq.\ \eqref{eq:graph} by a few alternatives. The first state we consider is simply a partially entangled state:
\begin{equation} \label{eq:partZ}
\ket{G^{\epsilon}_2}=\sqrt{\epsilon}\ket{0+}+\sqrt{(1-\epsilon)}\ket{1-}.
\end{equation}
Note that $\lvert G_{2}^{1/2} \rangle=\ket{G_2}$. Since we expect the one-way model to not be symmetric under arbitrary single-qubit rotations applied on $\rho_{in}$, we also consider rotated versions of $\ket{G^{\epsilon}_2}$, namely,
\begin{align} 
\ket{F^{\epsilon}_2} &= H \otimes \openone \ket{G^{\epsilon}_2}\label{eq:partXZ} \\
\ket{H^{\epsilon}_2} &= H \otimes H \ket{G^{\epsilon}_2}\label{eq:partX}.
\end{align}
where $H$ is the Hadamard gate. We also consider  a depolarized Bell state:
\begin{equation} \label{eq:depol}
\rho_{iso}=\epsilon \ketbra{G_2}{G_2} + \frac{(1-\epsilon)}{4} \openone.
\end{equation}
Note that all states described so far interpolate between separable and maximally entangled states as function of $\epsilon$.

Besides partially entangled states, we also consider states that exhibit only classical correlations. The goal is to determine how much causal influence, if any, can be achieved with only ``classical'' resources\footnote{It is a certain abuse of terminology to associate this case with classical resources, since the output of the one-way protocol is a quantum state.}. To that end, we consider the state
\begin{equation} \label{eq:diagZ}
\rho_{C}= \sum_{i,j=0,1} p_{ij} (\openone \otimes H)\ketbra{ij}{ij} (\openone \otimes H).
\end{equation}
As before, the corresponding rotated versions
\begin{align} 
\rho_{C'}&= (H \otimes \openone) \rho_{C} (H \otimes \openone), \\
\rho_{C''}&= (H \otimes H) \rho_{C} (H \otimes H).
\end{align}

For this set of input states we obtain the following values of $ACE_Q$:
\begin{align} 
ACE_Q(\ketbra{F^{\epsilon}_2}{F^{\epsilon}_2})&= \frac{4}{\pi}\sqrt{\epsilon(1-\epsilon)}  \\ 
ACE_Q(\ketbra{G^{\epsilon}_2}{G^{\epsilon}_2})&= 2\sqrt{\epsilon(1-\epsilon)}  \\
ACE_Q(\ketbra{H^{\epsilon}_2}{H^{\epsilon}_2})&= \frac{2}{\pi} E(1-2\epsilon) \\
ACE_Q(\rho_{iso})&=\epsilon \\
ACE_Q(\rho_{C})&=\frac{2}{\pi} |p_{00}+p_{11}-p_{01}-p_{10}| \\
ACE_Q(\rho_{C'})&= ACE_Q(\rho_{C''})=0
\end{align}
where $E(k)=\int_{0}^{\pi/2} \sqrt{1-k^2 \sin(x)^2} dx$ is the complete elliptic integral of the second kind.

\subsection{Upper bound on the causal influence of separable states in the one-way model}

In the main text we stated that any (pure) entangled state, when measured in a suitable basis, can display a higher $ACE_Q$ in the one-way model of quantum computation than any separable state. In order to prove this claim, we need the fact that there is a nontrivial upper bound for separable states, namely
\begin{equation}
    ACE_Q(\ketbra{+0}{+0})=\tfrac{2}{\pi}.
\end{equation}
We now present proof of this upper bound.

Suppose first that the shared resource state can be written as a convex combination of other states, i.e.\
\begin{equation}
\rho_{in}= \sum_i p_i \rho_i.
\end{equation}

Following Eq.\  \eqref{rhoBdo(a)} we can write the state $\rho_B(do(a))$ as
\begin{align}
\rho_B(do(a)) &= 
\nonumber \textrm{tr}_A\left(\Pi_{a} \sum_i p_i \rho_i \right) + X\left(\textrm{tr}_A\left(\Pi_{a^\perp} \sum_i p_i \rho_i  \right)\right)X \nonumber \\
&= \sum_i p_i \left[ \textrm{tr}_A(\Pi_{a} \rho_i) + X \left( \textrm{tr}_A(\Pi_{a^\perp} \rho_i) \right)X \right]. \notag\\
&= \sum_i p_i \rho_{B,i}(do(a)),
\end{align}
where $\rho_{B,i}(do(a))$ is the output state of second qubit, $B$, 
assuming that the first qubit was measured in basis $\{\ket{a},\ket{a^\perp}\}$ and that the shared state was $\rho_i$. Now we can write the $ACE_Q$, assuming the two intervention choices as $\ket{a}$ and $\ket{a^\perp}$, as 
\begin{equation}
ACE_Q(\rho_{in})= \underset{\ket{a}}{\mathbb{E}} TD(\rho_B(do(a)),\rho_B(do(a^\perp)) = \underset{\ket{a}}{\mathbb{E}} TD\left(\sum_i p_i \rho_{B,i}(do(a)),\sum_i p_i \rho_{B,i}(do(a^\perp))\right).
\end{equation}
Now recall that the trace distance is defined as 
\begin{equation}
TD(\rho,\sigma)= \frac{1}{2} \norm{\rho-\sigma}_1= \frac{1}{2} \textrm{Tr} \sqrt{(\rho-\sigma)^2}
\end{equation}
where $\norm{\cdot}_1$ is the trace norm. From this, we can write 
\begin{align}
\nonumber
ACE_Q(\rho_{in})&= \underset{\ket{a}}{\mathbb{E}} \norm{\sum_i p_i \rho_{B,i}(do(a)) - \sum_i p_i \rho_{B,i}(do(a^\perp))}_1  \\
&=\underset{\ket{a}}{\mathbb{E}} \norm{\sum_i p_i (\rho_{B,i}(do(a)) -  \rho_{B,i}(do(a^\perp))}_1.
\end{align}
Given that the trace norm is a norm, it is convex. Using also linearity of expectation we have that
\begin{align}
\nonumber
ACE_Q(\rho_{in})&=\underset{\ket{a}}{\mathbb{E}} \norm{\sum_i p_i (\rho_{B,i}(do(a)) -  \rho_{B,i}(do(a^\perp))}_1 \\ &\leq \sum_i p_i \underset{\ket{a}}{\mathbb{E}} \norm{(\rho_{B,i}(do(a)) -  \rho_{B,i}(do(a^\perp))}_1
\nonumber
\\
&= \sum_i p_i ACE_Q(\rho_i).
\end{align}
In other words, the $ACE_Q$, as a function of the shared resource state in the one-way protocol, is convex. Since any separable state $\rho_{sep}$ is a convex combination of product states, this means that no separable state can have an $ACE_Q$, in this context, higher than its component product states. Consequently, the largest $ACE_Q$ among all separable two-qubit states will be achieved by a product state.

Now suppose that the input state is an arbitrary two-qubit product state
\begin{equation}
    \rho_{prod}= \ketbra{\psi}{\psi} \otimes \ketbra{\varphi}{\varphi}.
\end{equation}
If we parameterize the two single-qubit states as
\begin{align}
\ket{\psi}=&\cos{\left(\tfrac{\theta_1}{2}\right)}\ket{0}+e^{i \phi_1}\sin{\left(\tfrac{\theta_1}{2}\right)}\ket{1}, \\
\ket{\varphi}=&\cos{\left(\tfrac{\theta_2}{2}\right)}\ket{0}+e^{i \phi_2}\sin{\left(\tfrac{\theta_2}{2}\right)}\ket{1},
\end{align}
a straightforward (if tedious) calculation shows us that
\begin{equation}
    ACE_Q(\rho_{prod})=\frac{2}{\pi} \sin{\theta_1} \sqrt{\cos{\theta_2}^2+\sin{\theta_2}^2\sin{\phi_2}^2}.
\end{equation}
Combining everything, we conclude that
\begin{equation}
    ACE_Q(\rho_{sep})\leq ACE_Q(\rho_{prod})\leq \frac{2}{\pi} = ACE_Q(\ketbra{+0}{+0}),
\end{equation}
as claimed.

\section{Causal influence in quantum teleportation - details}

We performed the same analysis as in the previous Section, but for the well-known quantum teleportation protocol. The behaviour is qualitatively similar in most aspects (unsurprisingly, since one-way quantum computation uses quantum teleportation as a primitive), so we will not repeat all arguments and proofs from the previous Section, instead focusing on the distinctions.

First, let us recall the ideal teleportation protocol. Two parties, Alice and Bob, share a Bell pair
\begin{equation}
    \ket{F_2}= \tfrac{1}{\sqrt{2}} (\ket{00}+\ket{11}) = H\otimes \openone \ket{G_2},
\end{equation}
and we label these qubits 2 and 3. Alice has another qubit, which we label 1, whose state she wants to teleport to Bob, and which we parameterize as follows 
\begin{equation}
    \ket{a}=\cos{\left(\tfrac{\theta_1}{2}\right)}\ket{0}+e^{i \phi_1}\sin{\left(\tfrac{\theta_1}{2}\right)}\ket{1}.
\end{equation}
Alice applies a $\cnot$ gate with qubit 1 (2) as control (target), followed by a $H$ gate on qubit 1. She then measures both of her qubits, and sends the measurement outcome to Bob. If the measurement outcomes of Alice`s qubits 1 and 2 were $s_1$ and $s_2$, respectively, for $s_i \in \{0,1\}$, Bob must apply a gate $Z^{s_1}X^{s_2}$ on his qubit 3, successfully recovering state $\ket{a}$.

We consider Alice`s choice of state to teleport, $\ket{a}$, as the intervention. In other words, we compute the $ACE_Q$ from Alice's choice to Bob's output state as 
\begin{equation}
ACE_Q(\rho_{in})= \underset{\ket{a}}{\mathbb{E}} TD(\rho_B(do(a)),\rho_B(do(a^\perp)),
\end{equation}
where $\rho_{in}$ is the shared two-qubit state, and
\begin{equation}
\rho_B(do(a))= \sum_{s_1, s_2} Z^{s_1}X^{s_2}\textrm{tr}_A\left(\Pi_{s_1, s_2} U (\ketbra{a}{a} \otimes \rho_{in}) U^\dagger \right) X^{s_2} Z^{s_1}.
\end{equation}
Here, $U =  H_2 \cdot \cnot_{12}$, $\Pi_{i,j}$ is the projector on outcome $i$ at qubit 1 and $j$ at qubit 2, and the partial trace is taken over both of Alice`s qubits. As before, we assume that the intervention consists of choosing between a pair of orthogonal states, $\ket{a}$ and $\lvert a^\perp \rangle$, and we average uniformly over all $\ket{a}$ (which now means averaging over the full Bloch sphere, not only the equator as in the previous Section).

\begin{figure}[h]
    \centering
    \includegraphics[width=0.6\textwidth]{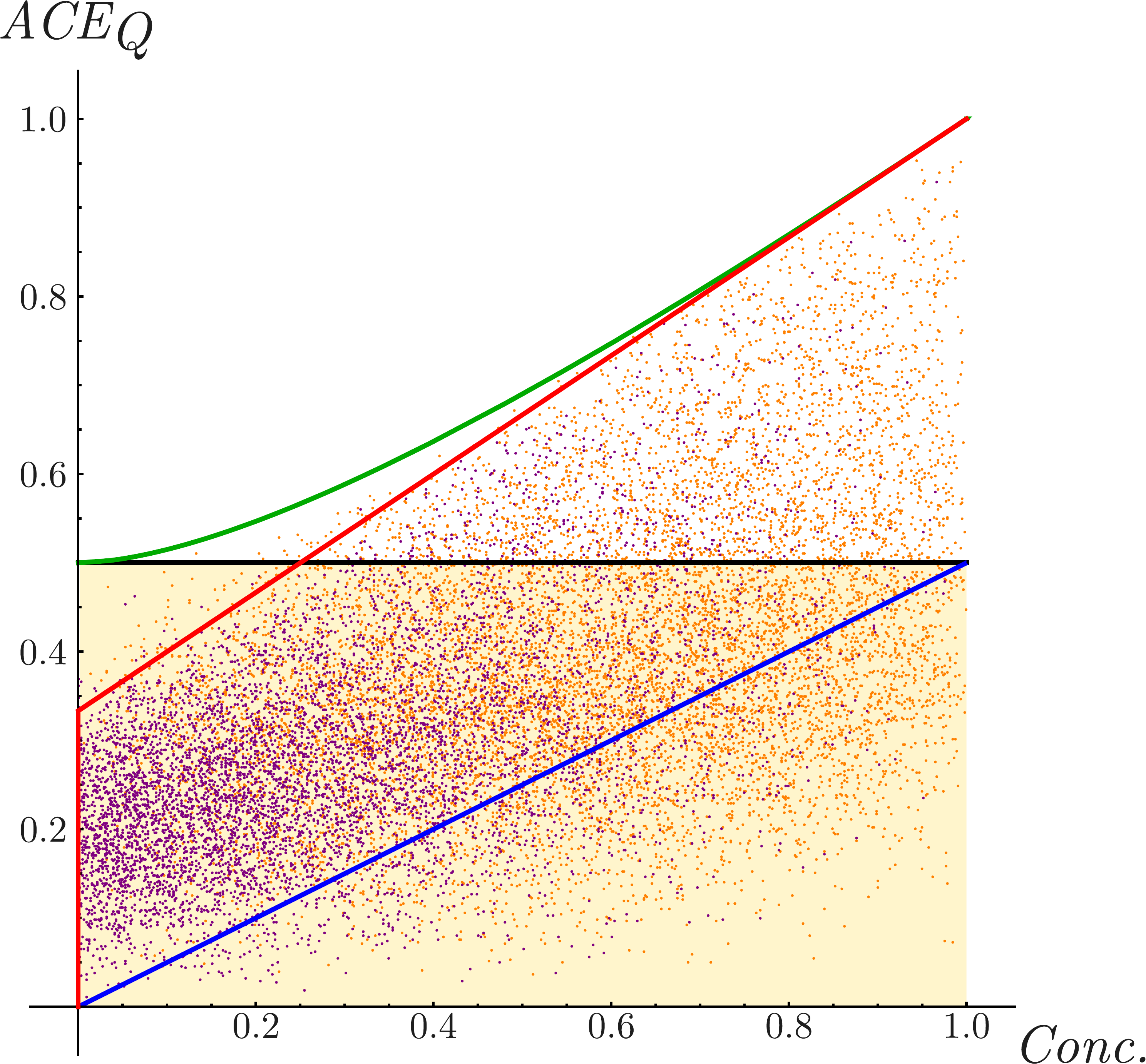}
    \caption{$ACE_Q$ as a function of the concurrence. The green and red curves correspond to families of states $\ket{F^{\epsilon}_2}$ and $\rho_{iso}$, respectively. The blue curve corresponds to either $\ket{H^{\epsilon}_2}$ or $\ket{G^{\epsilon}_2}$. The black horizontal line delimits the shaded region and corresponds to $ACE_Q(\rho_\textrm{sep}) = 1/2$. Orange points correspond to 10000 uniformly-sampled pure states, whereas purple points correspond to 10000 randomly sampled mixed states.}
    \label{figure:Teleportation}
\end{figure}

We expect there to be some nonzero causal influence from A to B even in the absence of entanglement, due to the classical communication that happens at the end of the protocol. To test that, we compute the $ACE_Q$ for the same families of quantum states shown in Eqs.\ \eqref{eq:partZ} to \eqref{eq:depol}. We were unable to obtain closed-form solutions for all cases, but the numerical results are shown in Fig.\ \ref{figure:Teleportation}. 

As anticipated, the plot of Fig.\  \ref{figure:Teleportation} is qualitatively similar to that of Fig.\ 2 in the main text. The most immediate differences are (i) less distinct behaviors among the families of quantum states, and (ii) the lack of an absolute lower bound for the value of the $ACE_Q$ for a given concurrence. Point (i) follows from the fact that the one-way protocol, as we described it, has a preferred direction in the Bloch sphere, since measurements are made only in a particular equator. This is why there is an important difference between states $\ket{G^{\epsilon}_2}$ and $\ket{H^{\epsilon}_2}$. Quantum teleportation, on the other hand, is an isotropic protocol, in the sense that everything should be basis invariant, and states that differ by a rotation of the type $U\otimes U$, for some single-qubit $U$, should display the same behavior.

Quantitatively, the main difference is that the upper bound for separable states is $1/2$ rather than $2/\pi$. However, it remains the case that any entangled two-qubit state, if measured in the correct basis, outperforms the best separable state, showing that quantum teleportation also displays a notion of quantum advantage over classical resources, at least from the point of view of causal influence.

\end{document}